\documentclass{aa}
\include{psfig}
\usepackage{graphics}
\begin{document}
\thesaurus{03.09.7;03.09.5; 03.13.5; 03.13.2;03.20.5; 09.18.1)}
\title{Acquisition and analysis of adaptive optics imaging polarimetry data}
\titlerunning{Adaptive optics polarimetry data analysis}
\author{N. Ageorges\inst{1,2} 
\and J. R. Walsh\inst{2}} 
\authorrunning{Ageorges \& Walsh}
\institute{
National University of Ireland - Galway, 
Physics Department, 
Galway,
Ireland. \\
E-mail: nancy@physics.ucg.ie
\and
European Southern Observatory, 
Karl Schwarzschild Strasse 2,
D-85748 Garching bei M\"unchen, 
Germany. \\
E-mail: jwalsh@eso.org}

\offprints{N. Ageorges}
\date{Received  / Accepted }
\maketitle

\begin{abstract}
The process of data taking, reduction and calibration of near-infrared 
imaging polarimetry data taken with the ESO Adaptive Optics System 
ADONIS is described. The ADONIS polarimetric facility is provided by 
a rotating wire grid polarizer. Images were taken at increments of
22.5$^\circ$ of polarizer rotation from 0 to 180$^\circ$, over-sampling
the polarization curve but allowing the effects of photometric variations to be assessed.
Several strategies to remove the detector signature are described. 
The instrumental polarization was determined, by observations of 
stars of negligible polarization, to be 1.7\% at J, H and K bands. The lack of
availability of unpolarized standard stars in the IR, in particular
which are not too bright as to saturate current IR detectors, is highlighted. 
The process of making polarization maps is
described. 
Experiments at restoring polarimetry
data, in order to reach diffraction limited polarization, are outlined, 
with particular reference to data on the Homunculus reflection
nebula around $\eta$ Carinae.
\end{abstract}

\section{Introduction}
High angular resolution techniques for astronomical imaging
have matured rapidly in recent years (see e.g. Beichman \& Ridgway \cite{beich}) and have
been applied to a variety of Galactic and extra-galactic
sources. For ground-based near-IR imaging, adaptive optics 
techniques have been very succesful in approaching the theoretical
diffraction limit of the telescope (Rigaut et al. \cite{rigaut}). 
The systems allow on-line 
correction for atmospheric perturbations using either a natural
guide star (part of the object under study, or an
unrelated star in the near vicinity) or laser guide star
(see e.g. Lloyd-Hart et al. \cite{lloyd}). 
A natural extension of this technique is to achieve 
two-dimensional polarimetric observations in the near-infrared
with the, in principle, simple provision of a polarizer in the 
beam. The combination of both techniques allows information on
the detailed polarization of an extended source, or determination 
of the individual polarization of close multiple sources. It may
be applied to reflection nebulae for determining the position of 
embedded illuminating sources, for study of the line of sight
geometry of dust scattering regions and for the orientation of
magnetic fields in star forming regions or quasar jets.

The extinction of interstellar dust peaks in the UV and 
declines to longer wavelengths (e.g. Mathis \cite{mathis}), but the continuum emission 
from grains at typical temperatures of a few hundred K in regions 
heated by starlight increases strongly above 2$\mu$m. In addition
molecular emission and absorption bands are stronger above 3$\mu$m. 
The 1-2$\mu$m region therefore provides an ideal window 
for the study of the close environment of dust embedded sources,
such as regions around proto-stars or emerging young stars.
For typical interstellar grains the low extinction in the near-IR enables 
information on the scattering properties of the grains, or the study of scattering 
regions, which have high optical extinction. 
Near-IR polarization is thus 
entirely analogous to optical polarization study but can be
extended to more embedded environments. At longer wavelengths 
the grain emission dominates and any polarization of the radiation 
is controlled by anisotropic emission mechanisms such as aligned 
non-spherical grains (Davis \& Greenstein \cite{dg}). 
Examples of IR polarimetry include: detection of extended dust disks in 
young stellar objects (see e.g. Piirola et al. \cite{piscalco}, for observational 
results and Berger \& M\'enard \cite{jeanphi}, for theoretical work); dust structures 
in AGB envelopes (e.g Sahai et al. \cite{sahai} for CRL 2688); detection of dust in 
interstellar jets (e.g. Hodapp 1984); magnetic field structure in star forming regions 
(e.g. Whittet et al. \cite{whitgera}) and polarization in galaxies 
(e.g. Jones \cite{jones}) and quasars (Sitko \& Yudong \cite{sitko}). 
Extending polarimetry to the IR also brings
the potential of high spatial resolution, both through the
dependence of diffraction on wavelength and the decrease in 
atmospheric seeing size with wavelength.  
 
  In the near-IR, the dominant contribution to polarization
is therefore from scattering of radiation by grains, 
and their finite relative size 
requires that Mie theory must be used to predict the scattering
properties. However the optical properties of typical interstellar
grains are fairly well represented by models based on laboratory
and observational data (Draine \& Lee \cite{draine}), so that
the scattering properties of interstellar grains in the near-IR can 
be predicted. Whilst polarization data naturally provides
geometric information on the location of illuminating sources,
the scattering efficiency with scattering angle is required to derive
geometric information about the line of sight location of the scatterers
(White et al. \cite{white}).
For high dust column optical depths, multiple
scattering may occur and has then to be modelled using Monte
Carlo methods (c.f. e.g. Witt \cite{witt}, Warren-Smith \cite{warren}; Whitney 
\& Hartmann \cite{whitney}; Fischer, Henning \& Yorke \cite{olaf}; Code \& 
Whitney \cite{code}). 

 As part of a programme to study the nature of the dust in the 
Homunculus nebula around the massive star (or stars) $\eta$
Carinae and determine information about the 3-D structure of 
the reflection nebula, near-IR imaging polarimetry data were
obtained with the ESO ADONIS system. $\eta$ Car and the Homunculus 
is an ideal source for adaptive optics since the
central point source is very bright and the nebula is not so
extended that off-axis anisoplanicity becomes an important effect.
The present paper is devoted to the details and subtleties of the
data collection and removal of the instrumental signature
vital to the derivation of a polarization map. A following
paper will present the scientific results on the high resolution
near-IR polarization of $\eta$ Car and the Homunculus. Sect. 2 
is devoted to a brief description of the ADONIS instrument;
Sect. 3 then considers the observational strategy. The
fundamentals of the data reduction are described in Sect. 4
and the polarimetric calibration of the instrument in Sect. 5.
Sect. 6 exposes the different deconvolution techniques applied to the data 
and their resulting effect on the polarization maps. 

\section{The ADONIS adaptive optics system}

\subsection{The adaptive optics system}
\begin{figure*}[htb]
\vspace{0.5cm}
\hspace{1cm}
\centerline{\resizebox{\hsize}{!}{\includegraphics{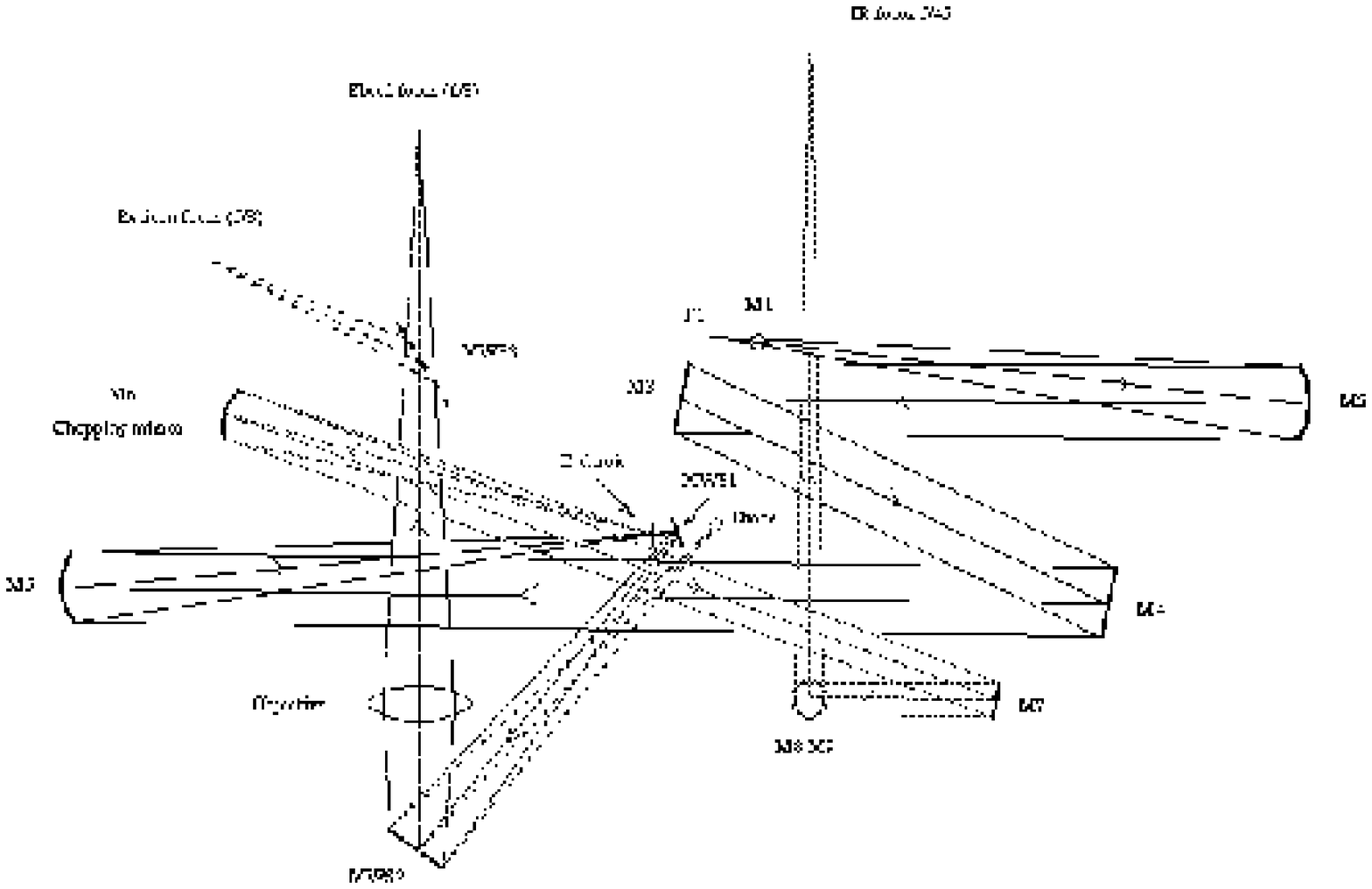}}}
\vspace{-0.0cm}
\caption{Optical layout of the ESO ADONIS adaptive optics system. 
The polarimeter, as a prefocal instrument, is installed at the
entrance window of the dewar (IR focus f/45).}
\label{layout}
\end{figure*}

ADONIS is the ADaptive Optics Near INfrared System (see e.g. Beuzit \& Hubin \cite{BH} 
or Beuzit et al. \cite{beuzit}) supported by ESO
for common users since December 1994 at the F/8.1 Cassegrain
focus of the La Silla 3.6 m telescope. Fig ~\ref{layout} 
shows the optical layout of the ADONIS adaptive optics (AO) 
system. A tip-tilt and a 64 element deformable mirror corrects the
distortions of the image in real time and a Shack-Hartmann wavefront
sensor (WFS) provides the difference signal for the deformable mirror using
a bright reference star close to the object. 
The detector for the Shack Hartmann sensor
can be chosen as either an intensified Reticon for bright sources
(m$_v$ $<$ 8 mag.) or an
electron bombarded CCD for fainter sources (8 $<$ m$_v$ $<$ 13 mag.,
25 to 200Hz sampling). Both detectors are sensitive in the visible 
wavelength region. 
An off-axis tiltable mirror allows the sky background, 
in a field of radius $\leq$30$''$, to be chopped with the on-source image.
The output F/45 focus delivers the image to a near-IR detector -
either a Rockwell 256$^{2}$ HgCdTe array (SHARP II for 1-2.5$\mu$m, Hofmann et 
al. \cite{hofmann}) or a LIR HgCdTe 128$^{2}$ anti-blooming CCD (COMIC for 1-5$\mu$m, 
Marco et al. \cite{marco}).  


\subsection{The camera}

The SHARP II camera was selected for the near-IR polarimetric
observations. This camera has a fast shutter 
at the internal cold Lyot stop, allowing integration times as short as
20msec. 
The present observations were made with the standard J, H, K 
filter set and a narrow band 2.15$\mu$m continuum filter, with a width of 
0.017$\mu$m, and denoted hereafter K$_c$. 


\subsection{The polarizer}
\label{polarizer}

The polarizer, from Graseby Inc.,  is a wire grid of 0.25$\mu$m period, 
on a CaF$_2$ subtrate. It is especially
designed to work in the spectral range 1 to 9$\mu$m and has a 
transmission of 83\% perpendicular to the wire grid at 1.5$\mu$m. 
It is remotely rotated 
by the ADOCAM control system to any desired absolute position angle
within tolerances of 0.1$^\circ$. 
This polarizer, as a pre-focal instrument, is inserted into the beam in 
front of the camera; and is not cooled. 
However since the polarizer is not
oriented perfectly perpendicular to the optical axis there is a small 
image motion on the detector when rotating the polarizer (see 
Sect.~\ref{derivpol} and Fig.~5).

\section{Observational technique}
\label{obsmode}

The magnification giving a pixel scale of 0.05$''$ has been selected to 
ensure an adequate sampling of the PSF, at H band. The field was thus
12.8$\times$12.8$''$; for the study of extended sources larger than
the field size it is obviously necessary to employ several
pointings and mosaic the resulting images after basic data reduction.
ADONIS has a 
limit of 30$''$ for the radial extent of the offset sky so values 
less than [30 - half detector size] ($''$) must be employed in order 
to have unvignetted sky background frames. 
Special care has been taken in the selection of the offset sky position to 
avoid any overlapping with the extended object observed. 
For all sources, object and chopped sky images were obtained at each 
position of the polarizer. 
A data cube of 256$\times$256 spatial pixels $\times$ M frames, where 
M is the number of object and sky frames, was acquired. 
Table \ref{tab-pol-standards} lists the details of the ADONIS 
polarimetry observations of the science and calibration sources. 

Sets of chopped images were obtained at nine different positions of the
polarizer, each 22.5$^{\circ}$ apart, from 0 to 180$^\circ$. The minimum
number of frames required to determine the linear polarization and its 
position angle is 3 (spanning more than 90$^{\circ}$ in position angle). By 
effectively 
oversampling the polarization curve (viz. the variation of detected 
signal with polarizer rotation angle) one can at least hope 
to average out shorter term variations in atmospheric transmission 
in order to improve the quality of the polarization measurement. 
Expressed in terms of the Stokes parameters I, Q and U
(see e.g. Azzam \& Bashara \cite{azzb}), I depends on the 
total signal whilst Q and U depend on the difference in signals between
images taken at polarizer angles of 0, 90, 45 and 135$^\circ$.
Then the linear polarization p is given by, ~~ 
$ p(\%) = 100 \times \sqrt(q^2 + u^2) $ ~~ 
where q=Q/I and u=U/I. The position angle of linear polarization is, ~~ 
$ \theta(^\circ) = 28.648 \times tan^{-1}(U/Q) $ ~~ 
(Serkowski \cite{serk}). Determining the polarization 
from $\geq$ twice as many images as necessary
leads to improvement in polarization accuracy provided that any
photometric variations are on timescales different from 
the exposure time of individual images at each polarizer angle.
The worst case scenario is when photometric variations occur on
a timescale similar to the exposure times, so that the
measured difference signals vary wildy - the polarization 
determined by fitting a cosine curve then approaches zero.
The chosen exposure times per polarizer angle were in the 
range 1 to 50 s depending on the source brightness (see Table
\ref{tab-pol-standards}). Observing a polarized source with the polarizer at 0 and
180$^{\circ}$ polarizer positions should give the same detected counts
and is therefore a direct way to monitor the photometric variations 
during the observational sequence. Column 8 of Table 
\ref{tab-pol-standards} lists half the difference (in percentage) between 
integrated counts in the star profile for the 0 and 
180$^\circ$ images (i.e. rms on the mean of the 0 and 180$^\circ$ signal
values). For R Monocerotis, the semi-stellar peak of the reflection
nebula NGC~2261, the aperture covers the central extended
source (full extent 8$''$), whilst for OH~0739-14, a reflection
nebula around an embedded young star, an area 10$''$
in size was used for the statistics. 

\begin{table*}
\begin{tabular}{lcccrrrl}
 Source & Type & Date & Band & No. & T$_{exp}$ & No. Poln. & 0-180$^\circ$ \\
        &      &      &      & Frms. & (ms)~~ & sequence  & semi-difference (\%) \\
\hline
HD~93737   & Low poln. & 1996 Mar 02 & K & 20 &  50 & 3 & 0.76,0.65,0.22 \\
           & standard. & 1996 Mar 03 & H & 20 &  50 & 2 & 2.32,1.01 \\
           &           & 1996 Mar 04 & J & 20 &  40 & 1 & 0.15 \\
\hline
HD~64299 & Low poln. & 1996 Mar 03 & J & 10 & 5000 & 1 & 0.11 \\
         & star &                  & H & 10 & 3000 & 1 & 1.01 \\
         &      &                  & K & 10 & 3000 & 1 & 0.89 \\
         &      & 1996 Mar 04      & J & 3 & 10000 & 1 & 0.05 \\
         &      &                  & H & 3 & 6000 & 1 & 0.05 \\
         &      &                  & K & 5 & 10000 & 1 & 0.65\\
\hline
HD~94510  & Low poln.  & 1996 Mar 04 & K$_c$ & 30 & 40000 & 1 & 0.32 \\
         & star &                  &  & &  & & \\
\hline
OH~0739-41 & Extended IR & 1996 Mar 02 & J & 4 & 30000 & 1 & 0.45 \\
           & poln. source &               & H & 4 &  5000 & 1 & 0.48 \\
           &         &                & K & 4 &  5000 & 1 & 0.20 \\
\hline
R Monocerotis & Extended IR & 1996 Mar 03 & J & 10 & 1000 & 1 & 1.26 \\
             &  poln. source &            & H & 20 & 400 & 1 & 0.24 \\
             &           &            & K & 30 & 100 & 1 & 0.18 \\
\hline
$\eta$ Carinae & Polarized & 1996 Mar 02 & K & 200 & 50 & 4 &  \\
           & source        & 1996 Mar 03 & H & 200 & 50 & 2 & \\
           &               &             & H & 100 & 50 & 1 & \\
           &               & 1996 Mar 04 & J & 200 & 50 & 2 & \\
           &               &             & K$_c$ & 100 & 50 & 2 & \\
\hline
\end{tabular}
\caption{List of polarization sources observed. The exposure time (T$_{exp}$) is 
given per frame.} 
\label{tab-pol-standards}
\end{table*}

\section{Data reduction}

The data reduction applied to AO polarimetry data consists of the 
removal of the detector signature and sky subtraction, which is
common to IR imaging in general, followed by registration and
derivation of the polarization parameters.

\subsection{Removal of the detector signature}
\label{flatfield}
The basic data reduction steps were performed with the `eclipse'
package (Devillard \cite{nico}). Flat-fields were acquired on the twilight sky at the 
beginning of each night of observation, in an exactly similar way 
as for the targets, at nine angles of the polarizer. The integration times 
were 7, 10 and 20s for the J, H and K bands respectively;
no flat field was taken with the K$_c$ filter. The flat field images
must first be processed to flag bad pixels, caused by either
permanently dead pixels or ones whose sensitivity undergoes large
fluctuation during the exposure. Two methods have been employed
depending on the number of frames available in a cube: 
sky variation or median threshold. 

The `sky variation' method works on a data cube, with preferably  
many planes ($^>_\sim$20) in order to obtain reliable
statistics on the variations. The standard deviation ($\sigma$) 
with frame number is computed for each pixel in the frame. 
A histogram plot of the standard deviations 
has a Gaussian shape representing the response to the, assumed 
constant, sky signal. All pixels whose response is too low (dead) 
or too high (noisy), compared to a central $\pm \sigma/2$ interval, 
are rejected. The `median threshold' method can be applied to
a small number of input frames (such as flat field data) and
detects the presence of spikes above or below the local mean in 
each individual image independently. If the signal is assumed to 
be smooth enough, bad pixels are found by computing the
difference between the image and its median filtered version, and
thresholding it. This latter method is not as stringent as using 
the temporal variation, but is the only possibility when there are
an insufficient number of images to calculate reliable statistics. 
Some bad pixels may however remain in the images after applying the 
bad pixel correction by either method; however the number is small 
and they can be manually added to the bad pixel map. 
Slightly different bad pixel maps were found for the different 
positions of the polarizer; which could be explained by a 
polarization sensitivity of the pixels ($\sim$1\%), since the NICMOS 
detector sensitivity is slightly polarization dependent, or simply by 
the random variation of hot pixels. 

Once corrected for the bad pixels, the twilight flats were 
normalized, then multiple exposures were averaged for the same 
position of the polarizer to derive the flat 
field maps. The target data cubes were corrected with the bad pixel 
map derived using the `sky variation' method from the background sky 
frames and divided by the flat-field to give flat-fielded, cleaned 
images, where the sky contribution is still to be subtracted. All
these operations were performed independently for the nine positions
of the polarizer. 

\subsection{Sky subtraction}
\label{skysub}
The sky background can be bright in the IR and may also be polarized so
it is criticical in the case of polarimetry to ensure that the
uncertainties introduced by sky subtraction are minimized.
Several tests were performed to determine the impact of the method
of sky subtraction, in conjunction with the bad pixel correction, on 
the data. 
The first method considers one sky and a bad pixel map for each 
position of the polarizer; the second method a single averaged sky 
(all polarizer positions confounded) but individual bad pixel maps for 
each position; whilst the third method uses the same averaged sky and 
bad pixel map for all polarizer angles. 

All three methods were tested (Ageorges \cite{nancy}) and the results 
demonstrated that the 
largest modification of pixel values, and therefore photometry, comes 
from the bad pixel map used. The third method produced the largest
discrepancies from the expected $cos(2 \theta)$ curve, where $\theta$ is
the polarizer position angle. 
The first method is clearly to be preferred
since the effect of any polarization of the sky signal on the target
data is correctly removed and any short term variation in sky
background is subtracted.

It was found, from sky background level in the polarization calibrator data, 
that the sky subtraction has been successful to better than 1\% 
(rms noise of 3.5 ADUs). 
For the 0 and 180$^{\circ}$ data, a further test of the quality of the 
sky subtraction was performed: the skies have been exchanged, i.e. 'sky 0' 
has been used for the data taken at PA 180$^{\circ}$ and conversely. This 
resulted in 'photometric' variations less than 0.05\%, thus giving us further 
confidence in our sky subtraction method.

\subsection{Photometric quality} 
\label{photometry}
\begin{figure*}[htb]
\vspace{-11.5cm}
\hspace{-2cm}
\leftline{\resizebox{\vsize}{!}{\includegraphics{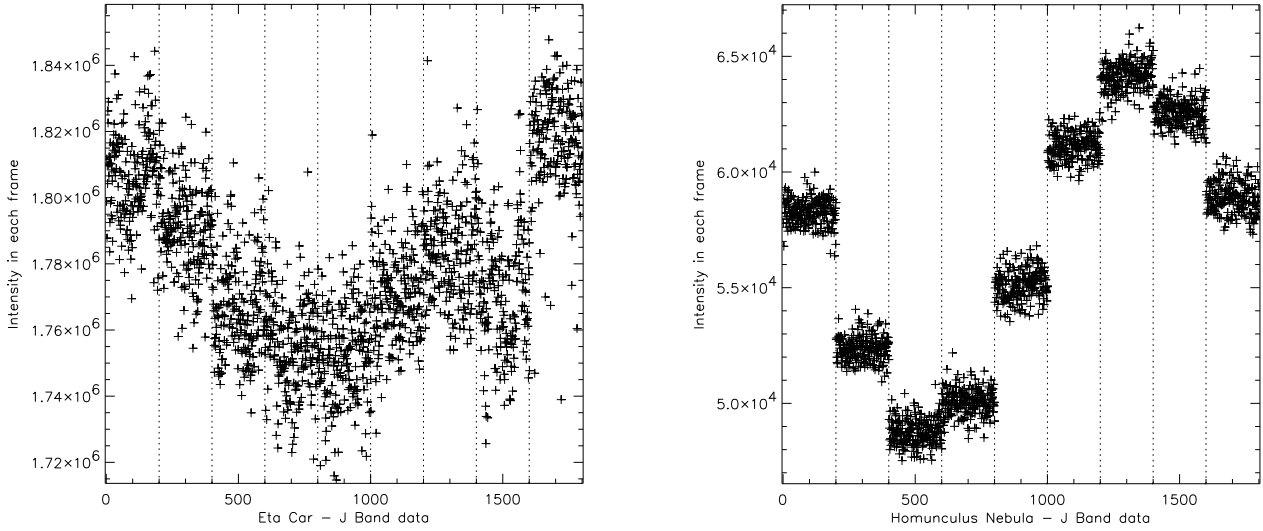}}}
\vspace{-12.5cm}
\caption{The photometric variation of the basic data is illustrated by the 
time sequence of measured counts in a region of the Homunculus
taken in J band. Every 200 frames, the polarizer has been rotated by
22.5$^{\circ}$ and the vertical dashed lines indicate the change of 
polarizer position angle. The width of this curve is characteristic 
of the photometric variations. The discrepant point at 
157.5$^{\circ}$ is attributable to an instrumental problem (see text). 
Left: global intensity variation over the full data frame but excluding 
a 30$\times$30 pixel box centered on $\eta$ Carinae; right: variation 
over a 50 $\times$ 50 pixel area of the nebula, far from the saturated 
center of the image. }
\label{new-intens}
\end{figure*}

The photometric quality of the data can be checked in two different
ways: either by comparing the photometry of an object when acquired at
0$^{\circ}$ and at 180$^{\circ}$ or by plotting the measured signal
against the polarizer angle where a $cos(2 \theta)$ form should be
obtained for polarized data. 
The latter is illustrated in Fig.~\ref{new-intens}, for J band data of the NE lobe 
of the Homunculus nebula around $\eta$ Carinae. The signal
is plotted with time as the polarizer was rotated from 0 to 180$^\circ$;
every ensemble of 200 points (within the dashed vertical lines)
corresponds to frames acquired at the same position of the 
polarizer. The spread of points at a given polarizer angle
gives a measure of the photometric variation. 

The images, used to create this plot, 
have been overexposed on purpose in order to get as much signal as 
possible on the faint nebula. The central region of the images has 
thus been obtained outside the linear regime of the CCD. 
The intensity variation over this image has thus been recalculated 
avoiding a 30$\times$30 pixels area centered on $\eta$ Car. This is 
represented Fig.~\ref{new-intens} together with a plot of the 
intensity variation over a 50$\times$ 50 pixels area centered on a 
lobe of the nebula, away from $\eta$ Car and thus obtained in the 
linear regime of the CCD. 
observed above is reduced by a factor of 2. 
Fig.~3, representing the  photometric variation of 
frames acquired at 0 and 180$^{\circ}$, clearly illustrates the fact 
that the night of these observations was not photometric: there is a 
0.3mag. extinction of the data acquired at 0$^{\circ}$ compared to 
that at 180$^{\circ}$.

In Fig.~\ref{new-intens} it is clear that there is
a discrepant point, at 157.5$^{\circ}$, since this does not fit into
the smooth $cos(2 \theta)$ progression of the curve. This problem, found 
for every source observed, was attributed
to a technical problem of unknown origin; it appears from
the figure that the polarizer may actually have been at an angle of 
45$^\circ$. All maps taken at this polarizer angle were ignored
in the subsequent derivation of polarization parameters, thus reducing the 
number of independent polarizer angles to 7 (0 and 180$^{\circ}$ being equivalent). 

\subsection{Derivation of polarization maps}
\label{derivpol}
The polarization degree for each pixel, binned pixel area or within 
an aperture was determined by fitting a $cos(2 \theta)$ curve 
to the variation of signal with polarizer rotation angle $\theta$ 
for the eight signal values (excluding the value at 157.5$^\circ$).
A least-squares procedure was used with linearization of the
fitting function and weighting by the inverse square of the 
errors (Bevington \cite{bev}). The error on the polarization 
was determined from
the inverted curvature matrix and the error on the position angle
by the classical expression (Serkowski \cite{serk}): ~~ 
$ \sigma_{\theta}(deg.) = 28.648 ( \sigma_{p}/p ) $ ~~ 
when $\sigma_{p}/p$ was $\geq$8 or from the error distribution of
$\sigma_{\theta}/\theta$ given by Naghizadeh-Khouei \& Clarke
(\cite{nag}) when $\sigma_{p}/p \leq 8$.
The errors on the individual points in the images at each polarizer
rotation angle take into account the number of images averaged, 
the read-out noise and the sky background contribution.
Since the detector offset is not fixed per image it was necessary to
bootstrap for the value of the sky level. 
A series of polarization maps were made with increasing sky contribution 
at a fixed polarization error per pixel. The sky signal was adopted when 
it produced polarization vectors which began to deviate from the expected 
centrosymmetric pattern (e.g. to the NE of R Mon - see Fig.~\ref{RMon}) 
in the regions of lowest signal. Thus the polarization errors are not 
absolute errors. Applying a polarization error cut-off to the maps 
produces maps consistent with the expected structure (which can also be 
partially checked by binning the data). 
Fig.~4
shows a typical fit to the $cos(2 \theta)$ curve for a 8$\times$8 pixels 
binned region of the R Monocerotis H band image (see Table 
\ref{tab-pol-standards} and Fig.~\ref{RMon}). 
The error bars on the individual
points arise from the photon statistics on the object and sky 
frames, with read-out noise considered.

It was noted in Sect.~\ref{polarizer} that the rotation of the 
polarizer induces an image shift on the detector. 
Fig.~5 
is an illustration of the displacement observed, for images of 
$\eta$ Carinae in K$_c$, while rotating the polarizer from 0$^{\circ}$ to
180$^{\circ}$ in steps of 22.5$^{\circ}$ (see Sect.~\ref{obsmode} for
details on the observation procedure).  Since the PSF is variable in
time, reproducability is not guaranteed. However the 
displacements were found to agree with those in Fig.~5 
for different targets (mostly unpolarized standard stars - see
 Table \ref{tab-pol-standards}), and in different filters, to better than 0.5 pixel 
and so were adopted to register the images at different polarizer 
angles.

For a point source, where only the integrated polarization is of
interest, the exact position of the source is not relevant provided 
all the signal is included in the summing aperture. However for
extended sources, such as for $\eta$ Carinae and the Homunculus nebula, 
a polarization map which exploits the available spatial resolution
is desired. It is therefore extremely important to ensure that the 
data are centered on the same position for all position angles 
observed, to avoid some smearing of the information. For unsaturated
stellar images, the centroid of the point source can be used as a 
fiducial to shift the images to a common centre. 
In the case of saturated images it proved possible to
obtain reliable centering by using a very large aperture for the
centroid; this is then weighted by the outer (unsaturated) regions
of the PSF. However if the source is polarized, and in particular
if there is polarization structure across the point source then
centroids at particular angles will be dependent on the source 
polarization. It was found that if the images were shifted
to match the centroids at the 8 polarizer angles for the R Mon
data, then a map with uniform, almost zero, polarization was
derived, in contradiction to the known (aperture) polarimetry
of this source (e.g. Minchin et al. \cite{minchin}). In such a case
the set of image shifts, derived from unpolarized point sources 
(Fig.~5), were applied to the data and the polarization 
maps were determined.
Fig.~\ref{RMon} shows the resulting J, H and K polarization 
vector maps superposed on logarithmic intensity plots; the raw
data has been binned 4$\times$4 pixels, i.e. 0.2$''$. 
Those shifts applied are closer to reality than those determined 
by the centroid of R Mon, but good to within $\pm$0.5 pixel. This 
might explain the difference in structure between our H band map 
and that of Close et al. (\cite{close}). 
extract of the Close et al. map. 
Considerable structure across the central (almost point) source 
is evident. The cut-off of the maps is determined by the value of 
the 1$\sigma$ polarization error (4, 4 and 6\% respectively for 
J, H and K).
The structures seen in the J, H \& K band maps (Fig.~\ref{RMon}) 
change with wavelength, which might be an optical depth effect 
of the inclined disk. 
The striking difference between the maps in Fig.~\ref{RMon} and 
the one reproduced in Ageorges \& Walsh (\cite{agewal}) comes from 
the calibration of the data. Indeed the latter were preliminary results 
and the first polarization maps derived with ADONIS. 

\section{Calibration of the polarimetric data}
\label{calib}

In order to determine the source intrinsic polarization and its position 
angle, several corrections are necessary. The instrument possesses
an instrumental polarization which must be vectorially subtracted
from the measured polarization. 
The instrumental contribution is derived from the observation of 
unpolarized standards. The interstellar medium between the source
and the observer also possesses an intrinsic polarization which needs to
be corrected. The typical ISM polarization values are $\le$2\% and can
be neglected when observing high polarization sources. If the
ISM polarization is not negligible, then it must be determined from
measurements of stars in the neighbourhood of the source (see e.g. 
Vrba et al. \cite{vrba}); alternatively the distance
dependence of the ISM polarization must be determined from
measurements of many stars. 
The zero point of the polarization
position angle is checked by observing non variable polarized standards, 
or polarized sources with reliable measurements. The lattest offer an
excellent check on the polarizing efficiency of the instrument
(i.e. response to a 100\% polarized source should be 100\%).

\subsection{Sky polarization}
\label{skypol}
In the optical during dark time the sky polarization
is typically 3-4\% (Scarrott, private communication). In the 
nights of our measurements, the sky polarization has been found 
to be consistent with zero within the error bars (typically
$\leq$0.5\%). 
Since it is the ratio of polarized intensity between the source and 
the sky that matters most and since the latter have carefully been 
subtracted (see Section~\ref{skysub}), the sky contribution has 
been ignored in processing the data. 

\subsection{Instrumental calibration}
\label{instcal}
\subsubsection{Choice of the polarization calibrators}
\label{polcal}

Despite extensive polarization observations, there 
is a distinct lack of any such standards in the IR. 
The polarized reflection nebulae OH~0739-41 
and NGC~2261 (illuminated by R Monocerotis) were
observed because of their extensive IR polarization data
(Heckert \& Zeilik \cite{heckert} and Shure et al. \cite{shure}
for OH~0739-14, Minchin et al. \cite{minchin} and Close et al. 
\cite{close} for R Mon), although neither can be claimed as true, 
non-varying standards. 

Since the observations are achieved using an adaptive optics 
system, the polarization standard could also be used as a PSF
calibrator. 
Since the correction is optimized continuously, the 
resulting PSF is variable in time. Any point source observed 
as PSF calibrator needs to be close ($<$ 10$^{\circ}$) to the 
target and be as similar as possible in terms of visible 
magnitude and spectral type, to ensure identical correction 
efficiency. Owing to the lack of polarization standards in the 
infrared, the polarization calibrators were chosen to be as close as 
possible to the source 
and bright enough to be used as reference for the wavefront sensor. 
In two cases, for HD~64299 and HD~95410, which have, respectively a B 
polarization of 0.151\% (Turnshek et al. \cite{turnshek}) and a V 
polarization of 
0.004\% (Tinbergen \cite{tinbergen}) it was assumed that the IR 
polarization is negligible, although no measurements exist at 
these wavelengths. 

In reducing the data taken on 1996 March 02 it was found that
the derived polarization for any source (even OH 0739-14) was consistent 
with zero polarization, and, in addition, did not exhibit the 
expected shift of image centroid with polarizer angle 
(Fig.~5. Either the photometric conditions were 
exceptionally poor (this is not borne out by large discrepancies 
between the 0 and 180$^\circ$ signal values - see Table 
\ref{tab-pol-standards}) or, more probably, an instrumental problem,
such as the polarizer not rotating to the requested angle, was
present. The polarization information
was therefore discarded for this night. However the K band
image of $\eta$ Carinae had excellent spatial resolution
and was retained (Walsh \& Ageorges, \cite{walage}).

\subsubsection{ADONIS instrumental polarization}
\label{instpol}

For the unpolarized (actually low polarization) standards, the integrated counts 
within a circular aperture including all the flux from the star profile 
(radius typically 
2$''$) above the sky background was measured for each angle of the
polarizer and a $cos(2 \theta)$ curve fitted to the data. 
Table \ref{pol-stan-data} lists the results. HD~93737 has a 
measured V band polarization of 1.07\% at position angle 122.4$^\circ$
(Mathewson \& Ford \cite{matfor}). Given the typical shape of
the interstellar extinction curve (the `Serkowski law', see
e.g. Whittet \cite{whittet}), the probable values of the
interstellar polarization for this star, assumed to have a typical
Galactic interstellar extinction, are
0.5, 0.3 and 0.2\% at J, H and K respectively. The position angle is
usually similar between the visible and IR (see eg. Whittet et al.
\cite{whitgera}). For the purposes of
computing the instrumental polarization it was assumed that the 
polarization was zero. 
The first two sets of data on HD~93737 on 
1996 Mar 02 (see Table \ref{tab-pol-standards}) are not
included on account of the problem with the data on that first 
night (see Sect.~\ref{instcal}). In addition the first sequence of
H band data on HD~93737 had poor photometry (see Table 
\ref{tab-pol-standards}) and was not considered. There is
a spread in the values indicating typical errors of $\pm$0.3\% in
linear polarization and $\pm$15$^\circ$ in position angle. 
Given the errors the J, H and K values are consistent with an instrumental
polarization of 1.7\%. 
Adopted values
are listed in the last row of the Table \ref{pol-stan-data}.
Given that only a single measurement was performed at K$_c$,
it is probably not significant that the instrumental polarization
in this band is higher and that the position angle differs from the K
band measurement.

\begin{table*}
\begin{tabular}{lccrrrr}
Target & Date & \multicolumn{1}{c}{J} & \multicolumn{1}{c}{H} & 
\multicolumn{1}{c}{K} & \multicolumn{1}{c}{K$_c$} \\
\      &      & \multicolumn{4}{c}{Linear poln. (\%) \& PA ($^\circ$)} \\
\hline
HD~93737 & 1996 Mar 02 &           & & 1.71, ~88 & \\
         & 1996 Mar 03 &           & 1.59, ~89 & & \\
HD~64299 & 1996 Mar 03 & 1.51, 111 & 1.99, ~86 & 2.16,  136 & \\
         & 1996 Mar 04 & 1.67, ~97 & 1.48, 112 & 1.71, ~89 & \\
HD~94510 & 1994 Mar 04 & 2.12, 104 &            &          & 2.05, 140 \\
\hline
Mean     &    -        &  1.74, 105 & 1.69, ~96 & 1.86, 104 & \\
Adopted  &     -       & 1.7, ~~105 & 1.7, ~~~90 & 1.7, ~~~90 & 2.0, ~~140 \\
\hline
\end{tabular}
\caption{Polarization of low polarization stars - instrumental polarization measurement}
\label{pol-stan-data}
\end{table*}

Once the instrumental polarization (intensity and angle) is determined
this correction can be applied to the polarization maps point-by-point.
Goodrich (\cite{goodrich}, in Appendix) describes the application of the 
instrumental correction.

\subsubsection{Position angle calibration}
\label{pacalib}

On producing polarization maps for the Homunculus nebula around 
$\eta$ Carinae, it was noticed that the polarization vectors
did not point back to the position of $\eta$ Carinae. There is
no reason for such a behaviour 
since it is known to be a reflection nebula. If the illumination
were by an extended source then the offset should not be one of simple
rotation. 
A novel method was
used to determine the single offset required to aligned all the 
polarization vectors in a centrosymmetric pattern around the position
of $\eta$ Carinae. A least squares problem was solved to minimize 
the impact parameter at the position of $\eta$ Carinae produced by the 
perpendiculars to all the 
polarization vectors in the Homunculus by application of a single
rotation. A consistent value of 18$\pm$1$^\circ$ was found for the
J, H and K images. In order to verify that this was not an artifact
of the $\eta$ Carinae nebula and the fact that the central point source
was saturated, the 18$^\circ$ correction was applied to the
polarization maps of NGC~2261. It was found that the vectors
in the high polarization spur to the NE were well aligned with the 
direction expected for illumination by the peak of R Mon. 
Thus the calibration of the absolute position angle can be made
without reference to a polarized standard. 

\begin{table*}
\begin{tabular}{lrrr}
Data source & \multicolumn{3}{c}{Polarization (\%) \& PA ($^\circ$)} \\
            & J~~~~~ & H~~~~~ & K~~~~~ \\
\hline
This work (PA uncorrected)    &  10.6, ~77  & 11.1, ~74 & 8.1, ~77 \\
Minchin et al. \cite{minchin} &  11.1, 100  & ~8.5, 103 & 5.6, 102 \\
\hline
\end{tabular}
\caption{JHK Polarization of R Monocerotis in an 8$''$ aperture}
\label{tab-rmon-pol}
\end{table*}

\section{Results on restoration of polarization images}
\label{restore}

In order to measure polarization structure in the vicinity of a 
bright point source, it is necessary to deconvolve the point 
source response from the data frames taken at each position
angle of the polarizer and then to form the polarization maps from 
the deconvolved images. The aim here is to detect polarization structure 
within an offset distance of a few times the diffraction limit from 
the point source. Several different approaches to 
restoration have been attempted in order to obtain detailed 
information on the fine structure of the Homunculus nebula  
close to the central source $\eta$ Carinae. This was motivated 
by the need to detect and measure the polarization of the three
knots found in the 0.4$''$ vicinity of $\eta$ Car by 
speckle imaging in the optical (Weigelt \& Ebersberger \cite{weiebe}
and Falcke et al. \cite{falcke}). The polarization data for
$\eta$ Car will be used to exemplify these experiments;
the scientific conclusions will be reported in Walsh \&
Ageorges (\cite{walage}). A preliminary discussion of restoration
of these images, without considering the polarization, has been
given by Ageorges \& Walsh (\cite{agewalspie}).

\subsection{Image restoration trial}

Two deconvolution techniques have been applied to the data: 
Richardson-Lucy (R-L) iterative deconvolution (Lucy \cite{lucy}, 
Richardson \cite{richard}) and
blind deconvolution (`IDAC', Jefferies \& Christou \cite{jeff}, 
Christou et al. \cite{christou1}). The major difference 
between these methods is related to 
the treatment of the point spread function (PSF). With the
Richardson-Lucy method, a PSF is required a priori to deconvolve 
the data, while for blind deconvolution, the PSF is determined from 
variations in the target object data. The blind deconvolution method 
uses an initial estimate, which can be a Gaussian for example. Since 
the adaptive optics PSF changes with 
time and is not spatially invariant (see e.g. Christou et al. 
\cite{christou2}) , blind deconvolution should be
better suited than the Richardson-Lucy method, which assumes a 
PSF constant in time. 
The exact spatial variation of the AO PSF is not
known. However in the present case, this is a minor problem since the
source itself ($\eta$ Car) has been used as wavefront sensor
reference star. Moreover with the pixel scale chosen, all the valuable
information in the short band data is enclosed in the isoplanatic
angle; the spatial variation of the PSF is thus negligible over the
area of the $\eta$ Carinae images, which is
not the case for the time variation. 

A comparison of the Richardson-Lucy method and IDAC - 'Iterative 
Deconvolution Algorithm in C', i.e the blind deconvolution algorithm used, 
was made using the K$_c$ data on $\eta$ Car (Table 2). The aim was to
test the reality of structures revealed in the near environment
of the central star of this reflection nebula. For the R-L
restoration, the Lucy-Hook algorithm (Hook \& Lucy 
\cite{hook}), in its software implementation under IRAF (`plucy'), 
was employed. The principle is the same as for the Richardson-Lucy
method, except that it restores in two channels, one for the
point source and the other for the background (considered smooth
at some spatial scale). The estimated position of the point source 
is provided and the initial guess for the background is flat.
K$_c$ data taken at polarizer angles of 0 and 180$^\circ$ were
restored (called K$_c$0 and K$_c$180). For the K$_c$0 image, blind
deconvolution was also performed.
It should be noted that although the polarizer angles are
effectively identical, the Strehl ratio is not identical between
the two data sets (K$_c^1$ \& K$_c^2$) and is higher for K$_c^1$ 
(27.9\% against 22.1\%). Although this could
be considered an advantage, it has a drawback since the four bumps
around the PSF (see Fig.~\ref{psf-ima} for the appearance of the
PSF) are more pronounced. These bumps (`waffle pattern') correspond to a null 
mode of the wavefront sensor as a result of an inadequacy in the control loop. 
The problem of the four bumps distributed symetrically 
around the source is that although they are in the
PSF they do not vary; they are fixed in time and position and 
therefore not removed from the image as part of the PSF. There is 
however a way to overcome this problem, and that is by forcing them 
to be in the PSF. 

Fig.~\ref{deconv} presents the deconvolution 
results obtained with both methods on the two separate data sets 
(Table \ref{tab-pol-standards}) and Fig.~\ref{psf-ima} shows the PSF derived 
from blind deconvolution. 
The 'plucy' deconvolved data have been restored to
convergence and then convolved with a Gaussian of 3 pixels FWHM.
The blind deconvolved data were not restored to convergence but limited to 1000 iterations 
to be comparable, in terms of number of iterations, with the Lucy deconvolution. 
The resulting image seems thus more noisy than the Lucy deconvolved ones. 
Note that neither of the methods used succeeded in removing the 4 bumps from the 
K$_c$ images of the first observational sequence (K$_c^1$0).

The data acquired at the polarizer position angles of 0 and 
180$^{\circ}$, deconvolved with the same algorithm 
(`plucy') both show identical structures (upper row of 
Fig.~\ref{deconv}). 
This example serves to illustrate the stability of 
the `plucy' method when applied to AO data while using a reasonable 
PSF estimate. The image 
from the first polarizer sequence, polarizer angle 0$^\circ$, 
deconvolved using IDAC is shown as the lower right image in 
Fig.~\ref{deconv} and is to be compared with the upper left image
deconvolved with `plucy'. It is clear that similar structures 
appear in both restorations and that there are no significant 
features in one restoration which do not appear in the other. 
The differences in the images are 
mainly due to the fact that the blind deconvolution has been 
stopped before fully resolving the data and the final image 
is thus more noisy. 
Moreover the presence of the four bumps is enhanced 
in this image. The major difficulty in this deconvolution 
is that these noise structures are convolved with extended 
emission from the Homunculus nebula. 
Being in the middle of the nebula,
the flux identified on these bumps is then a convolved product of the
waffle pattern  and the extended structure of the nebula. It is
thus very difficult for the program to isolate these four 'point
sources' and recover properly the true shape of the nebula at these
positions. 
In order to fully compare the different deconvolution techniques, blind 
deconvolution has been pushed to convergence for K$_c$0 (data set 1 \& 2). 
The results (Fig.~\ref{rec-ima}) are to be compared with the right hand side 
of Fig.~\ref{deconv}. 
The structures close to $\eta$ Car emphasized by the two
deconvolution processes, excluding the four bumps, 
confer a degree of confidence in the scientific results 
which will be presented in Walsh \& Ageorges (\cite {walage}).

\subsection{Polarimetry restoration trial}

In the case of polarimetric data, the deconvolution problem is more 
severe since the photometry must be preserved in the restored images
in order to derive a polarization map. The Richardson-Lucy algorithm 
is superior to blind deconvolution in that it should preserve flux. 
Experiments were performed on the K$_c$ $\eta$ Car data set,
restoring each of the nine polarizer images with the PSF derived 
from the unpolarized standard at the same polarizer 
angle. The results were poor even when the restored image
was convolved with a Gaussian of 3 pixels FWHM. They illustrate
the effect of the variable PSF and thus the difficulty to
recover polarization data at high angular resolution so close to the
star. 
Huge fluctuations in the
value of the polarization were seen in the vicinity of $\eta$ Car.
The differing PSF of the unpolarized star and of $\eta$ Car (the
AO correction was much better for the $\eta$ Car images than for
the standard star) produced restored images with large differences
in flux at a given pixel in the different polarization images.
At present there is no known method to recover the true PSF from the 
data and conserve the flux through restoration. A possible 
(although computer intensive) solution is to determine the PSF 
from blind deconvolution and use the result for the PSF 
in another algorithm known to preserve the flux. 
This has been performed here: the PSF determined by blind deconvolution has 
been used both with the Richardson-Lucy and Lucy-Hook algorithms. 
Since the IDAC blind deconvolution algorithm normalised the input 
image at the beginning of the iterations, the final image was rescaled back 
to the original total count to allow error estimation of the polarization image. 
Polarization maps for the three methods (`IDAC' alone and combined with R-L 
and `plucy' methods) 
have been created and compared after reconvolution with a 3 pixel Gaussian. 

From the high resolution restored images an attempt has been made to
derive the polarization map. Fig.~\ref{highp} illustrates the result obtained 
while using the PSF determined by the blind deconvolution with the R-L algorithm 
(30 iterations with the accelerated version), after reconvolution with a 
Gaussian of 3 pixel FWHM. The overall centro-symmetric pattern of 
polarization observed at larger scale and resolution is recognisable here as 
well. The major deviation from this pattern at $\Delta \alpha$ and 
$\Delta \delta$ zero (i.e. east-west and north-south through the image of 
$\eta$ Car) is due to the spider of the telescope. The presence 
of this feature is hard to identify on the intensity map underplotted but 
clearly present at this position in the original (undeconvolved) data. 
Fig.~\ref{highpcomp} is a vectorial difference between results obtained with Lucy 
deconvolution and blind deconvolution. 
Special care has been taken to avoid the vector difference to add when the position 
angles were separated by close to 2$\pi$. 
Some vectors at the border of the noise cut-off (e.g. at $\Delta 
\alpha \approx -3.0''$) detected in the Lucy map but not in the other are 
not represented here to avoid confusion with the differential vectors plotted. 
Major differences can be found at approximately 0.5$''$ from the center and 
correspond to differences in the deconvolution due to the wings of $\eta$ 
Carinae. 
At $\Delta \alpha$ = 0 and $\Delta \delta$ = 0 $\pm$ 0.3$''$, the
important difference between the two reconstructed polarization maps is
meaningless since these positions correspond to the spider of the
telescope and the data are poorly restored here. 



\section{Conclusions}

The process of data acquisition and reduction for polarization observations 
taken with the ESO ADONIS adaptive optics system has been described. Whilst
certain precautions both in the observing method and in data reduction
are required for imaging polarimetry and adaptive optics seperately, several
other problems are presented arising from the combination of the two methods. 
\begin{itemize}

\item Since the PSF varies in time the wavefront sensor reference star should 
be as similar as possible to the target object in terms of brightness
(since the achieved Strehl ratio strongly depends on the reference star
magnitude) and spectral characteristic, to ensure similar AO correction. 
Since the PSF varies across the field of view (depending on the anisoplanatic 
angle), it is also preferable to select the WFS reference star as close as 
possible to the target object. In practice this is rarely achieved when the 
target itself can not be used as WFS reference star. 
However a good estimate of the PSF 
provided by the reference star allows accurate deconvolution of the
target without the introduction of artifacts arising from the differing
PSF's.
\item Imaging polarimetry requires good photometric conditions. 
By oversampling the cos(2$\theta$) 
polarization curve at more than three position angles of the polarization 
analyser, an averaging over the photometric conditions is achieved.
However depending on the time period of the photometric variations 
the averaging can result in zero polarization even from a substantially
polarized source. 
In principle the use of an AO system should not compromise the photometric 
quality of the observations. 
\item The two polarization calibrations that are required impose the 
observation of an unpolarized source, to determine the instrumental
polarization, and of a target with known polarization, to calibrate the
angle of polarization. 
In the IR there is a very distinct lack of 
unpolarized and polarized standards. Stars with known very low
optical polarization are suitable as IR unpolarized standards since the
Serkowski interstellar polarization law shows that the polarization
is much less in the IR than the optical. However polarized standards
typically have a circumstellar origin to their high polarization and
the value in the IR cannot be predicted. Many of the reflection nebulae around 
Young Stellar Objects have variable polarization and are therefore 
not ideal polarized standards.
\end{itemize}
Several strategies have been described for flat fielding and sky subtraction
and it was shown how deviations from the expected cos(2$\theta$) curve can give 
an indication of the photometric conditions at the time of observation
and allow any discrepant polarizer angles to be discarded as was found
for the ADONIS polarizer at PA 157.5$^\circ$. The instrumental polarization
for ADONIS was determined at 1.7\% over the J, H and K range. Polarization
maps have been succesfully produced for the reflection nebula around 
$\eta$ Carinae (the Homunculus). By
using the PSF's determined from blind deconvolution at the same polarizer
angles as the data, it has been shown that polarization structure can be
revealed as close as two times the diffraction limit to a point source.
The interpretation of the ADONIS AO polarization results on $\eta$ Carinae 
will be presented in a forthcoming paper (Walsh \& Ageorges \cite{walage}).

\section{Acknowledgements}

We would like to thank the ESO ADONIS team for their advice and 
help during the development of the data reduction strategy.
S. M. Scarrott is also acknowledged for useful comments on 
imaging polarimetry.

{}

\begin{figure*}[htb]
\vspace{-11.5cm}
\hspace{-2cm}
\leftline{\resizebox{\vsize}{!}{\includegraphics{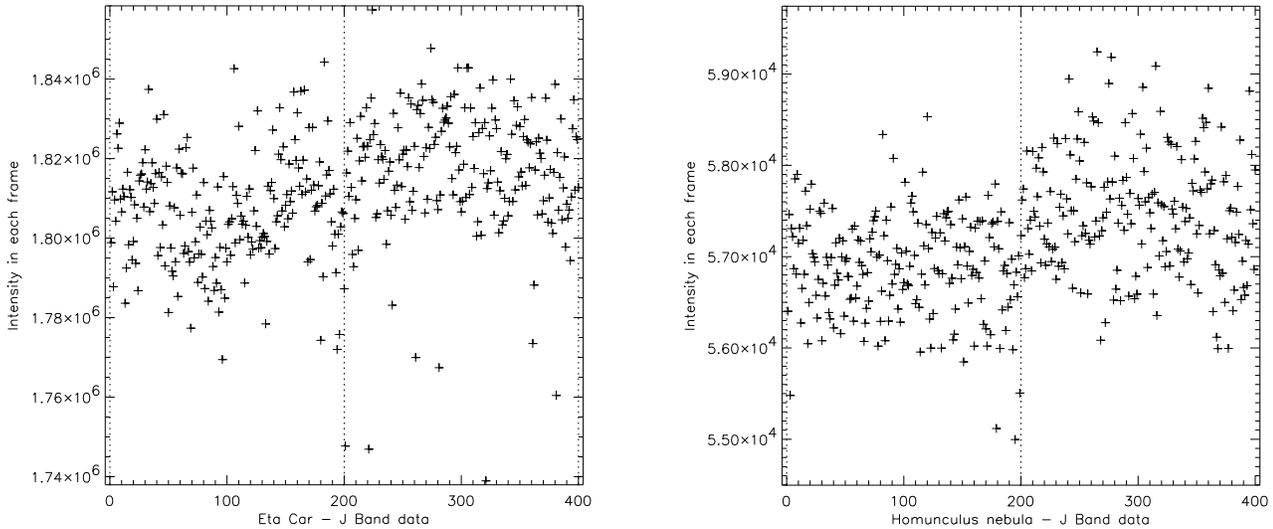}}}
\vspace{-12.5cm}
\caption{Photometric variation between data taken at PA 0 and 
180$^{\circ}$ is shown for the full frame (excluding the central 
source, i.e. $\eta$ Car itself) at left, and for a small area 
centered on the Homunculus (right).}
\label{new-int}
\end{figure*}

\begin{figure*}[htb]
\vspace{-4cm}
\centerline{\resizebox{\vsize}{!}{\includegraphics{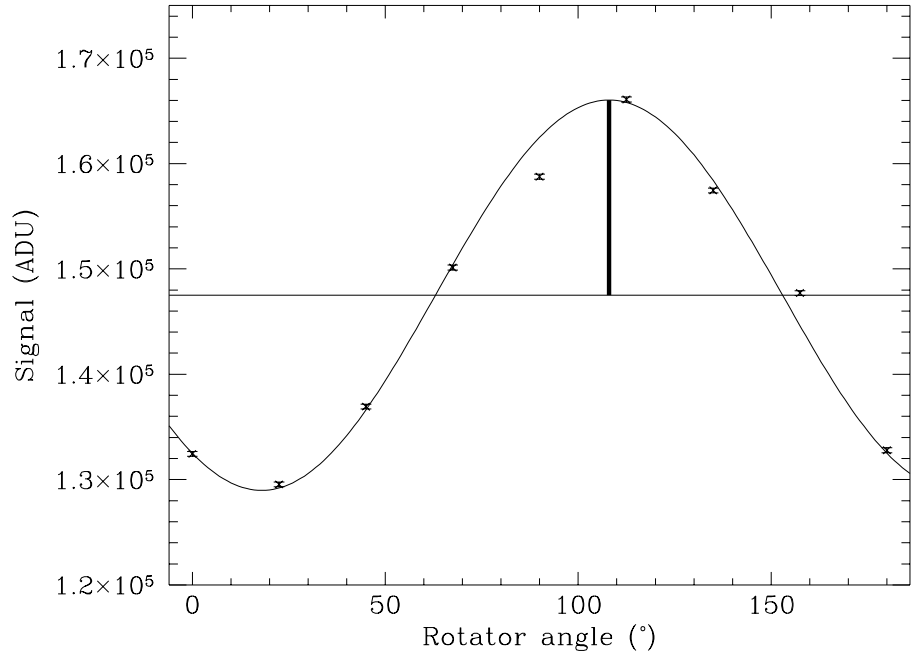}}}
\vspace{-16.5cm}
\caption{A typical fit of the observed signal as a function of
polarizer rotator angle by $p cos(2 \theta)$ for the summed counts
in an aperture over the H band image of R Monocerotis
(see Table \ref{tab-pol-standards} and Figure \ref{RMon}). 
The derived value of linear polarization and position angle is 
shown by the bold line. The point at 157.5$^\circ$ was not
considered in the fit.}
\label{coscurve}
\end{figure*}

\begin{figure*}[htb]
\centerline{\resizebox{\hsize}{!}{\includegraphics{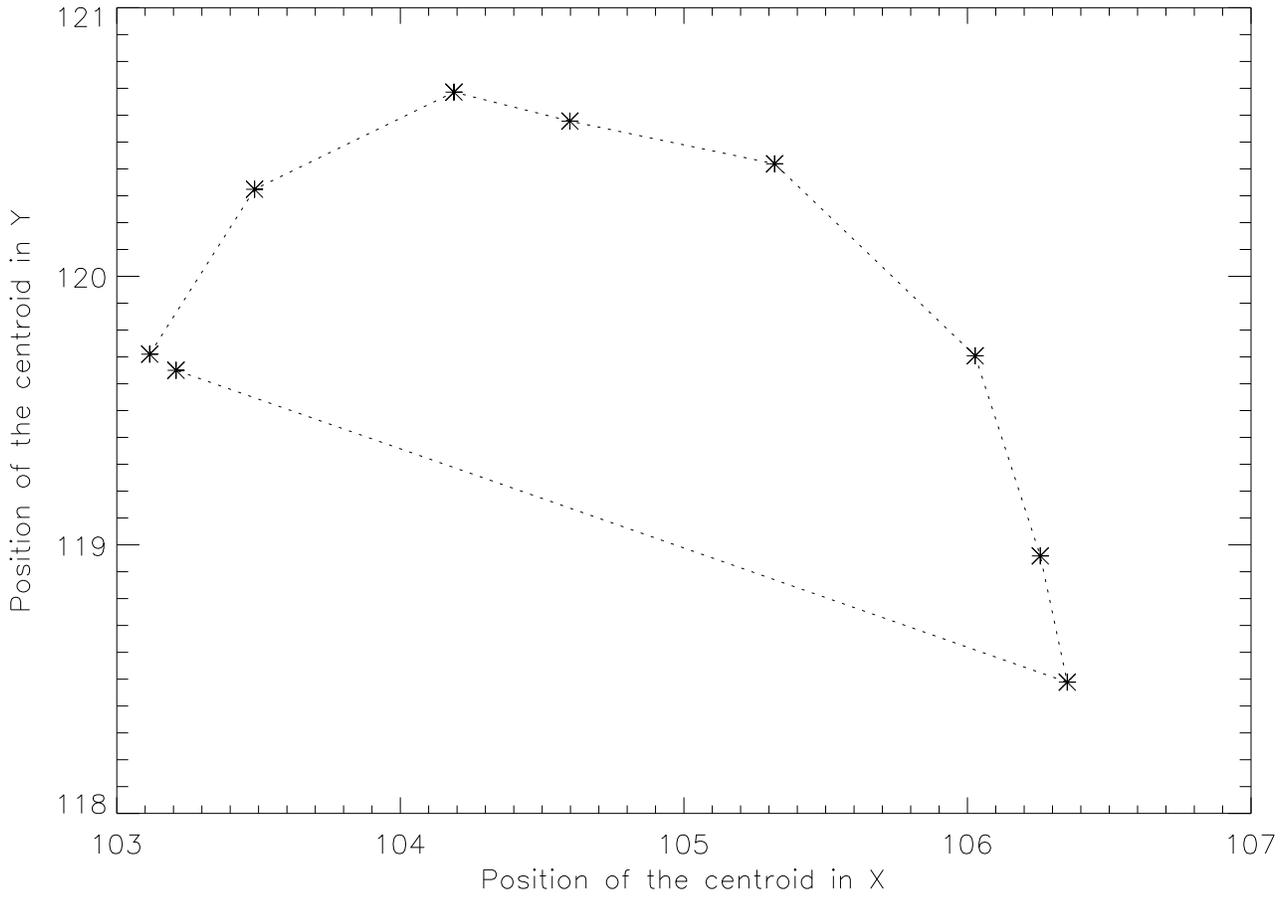}}}
\caption{Illustration of the variation of the position of the centroid
of an image on the detector while rotating the polarizer from 0 to 
180$^\circ$ in steps of 22.5$^\circ$. The target was $\eta$ Car observed 
in K$_c$ band and the shifts are clockwise with increasing rotator angle.}
\label{shift}
\end{figure*}

\begin{figure*}[htb]
\vspace{-13cm}
\hspace{1cm}
\centerline{\resizebox{\vsize}{!}{\includegraphics{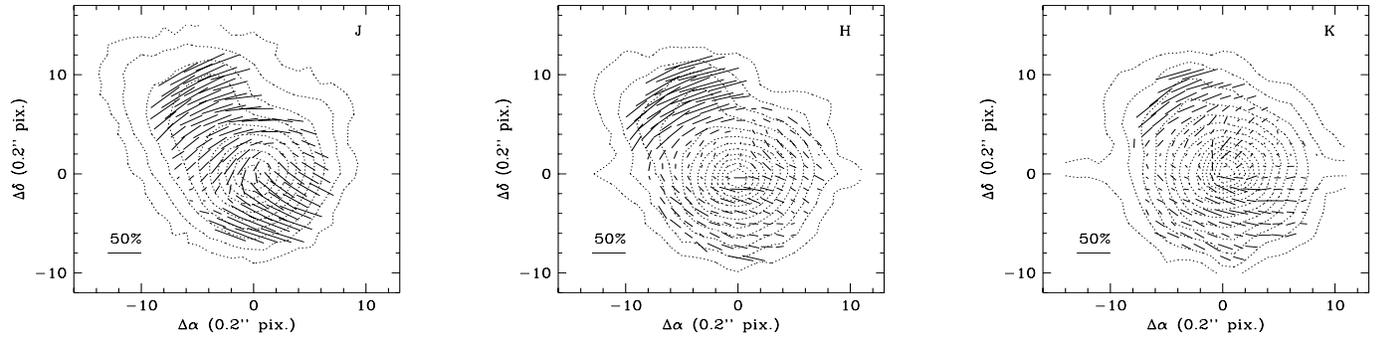}}}
\vspace{-13cm}
\caption{J, H and K polarization maps of R Monocerotis and the brightest
regions of the reflection nebulae NGC~2261. The contour maps shows the
logarithm of the signal (polarized + unpolarized). The size of the
polarization vectors is indicated and the orientation is north to the top
and east left.}
\label{RMon}
\end{figure*}

\begin{figure*}[htb]
\vspace{-7.0cm}
\centerline{\resizebox{\vsize}{!}{\includegraphics{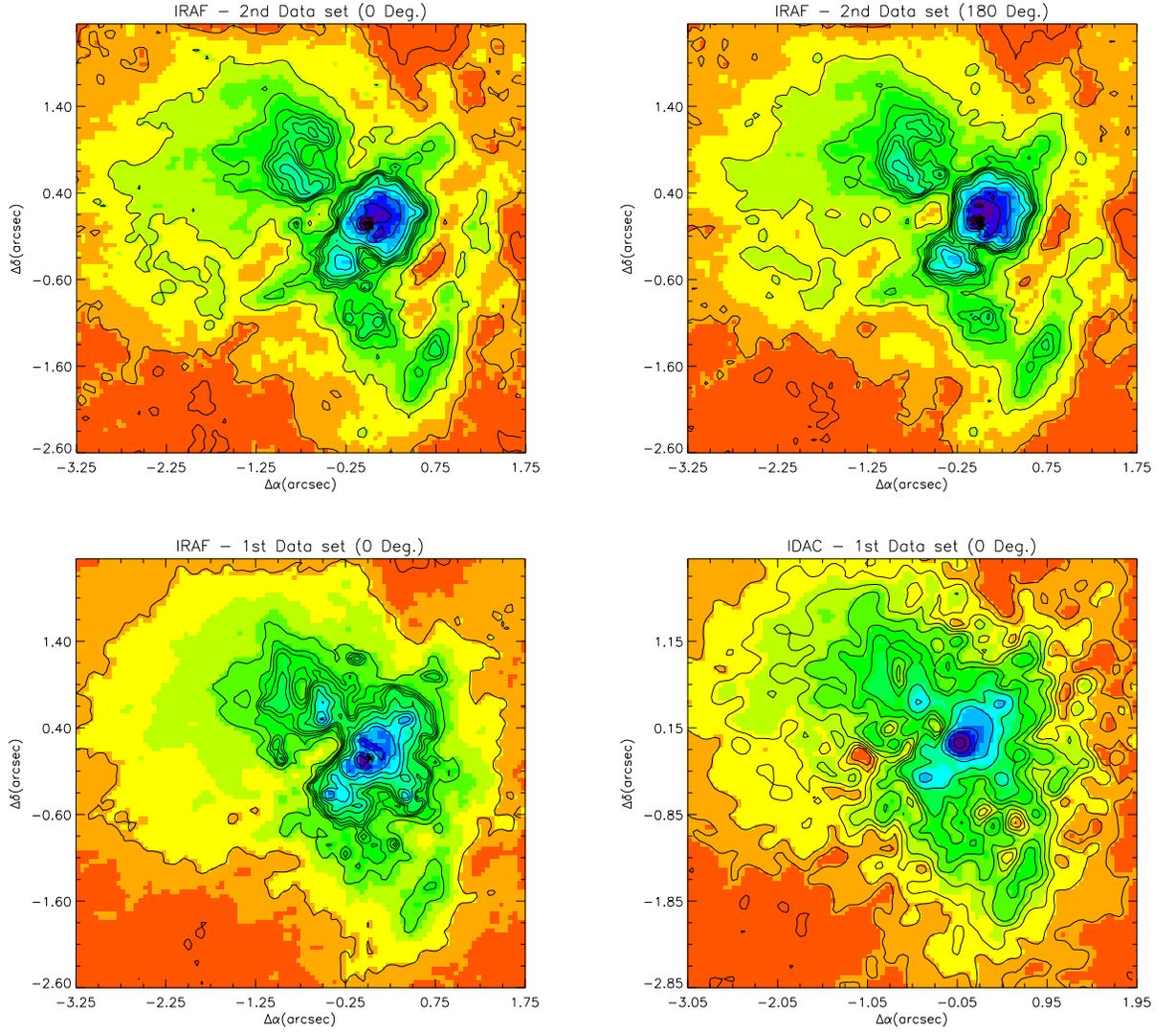}}}
\vspace{-10.5cm}
\caption{Comparison of the results of two deconvolution techniques 
for K$_c$ data of $\eta$ Carinae and the Homunculus nebula. On the
top row are shown data from the second polarizer sequence, 
at 0 (left) and 180$^\circ$ (right), both deconvolved with 
`plucy' and convolved with a Gaussian of 3 pixels FWHM. The bottom 
left image is identical to the top left one but the data restored
is from the first polarizer sequence (0$^\circ$ image). The result 
of blind deconvolution is presented in the bottom right hand image 
for the first data set - K$_c^1$ - at 0$^{\circ}$ polarizer angle.}
\label{deconv}
\end{figure*}

\begin{figure*}[htb]
\vspace{-1.0cm}
\centerline{\resizebox{\hsize}{!}{\includegraphics{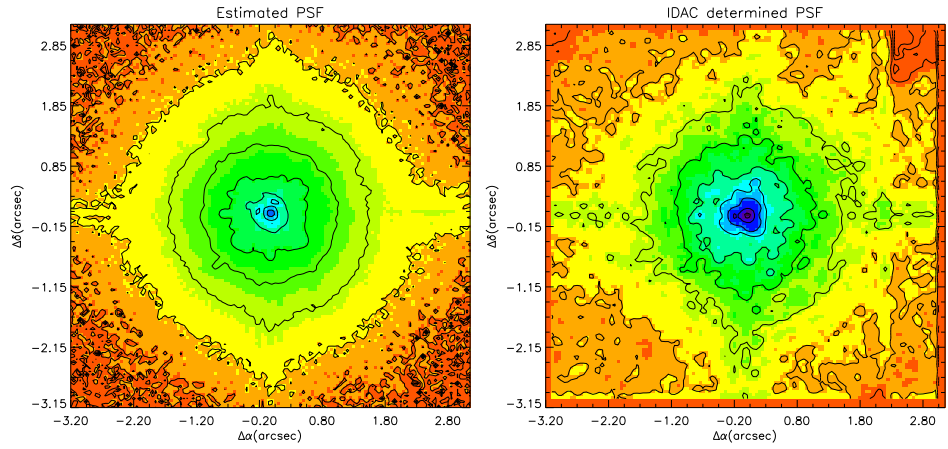}}}
\vspace{-9.0cm}
\caption{PSF estimated (left) and obtained (right) from an IDAC blind 
deconvolution of the K$_c$ image of $\eta$ Car (see Table 2 for 
observational details). The PSF estimate is the shift-and-add of 
the point source (HD 94510) observed shortly before the target 
data with the K$_c$ filter. It has been used as first guess for 
blind deconvolution.}
\label{psf-ima}
\end{figure*}

\begin{figure*}[htb]
\centerline{\resizebox{\hsize}{!}{\includegraphics{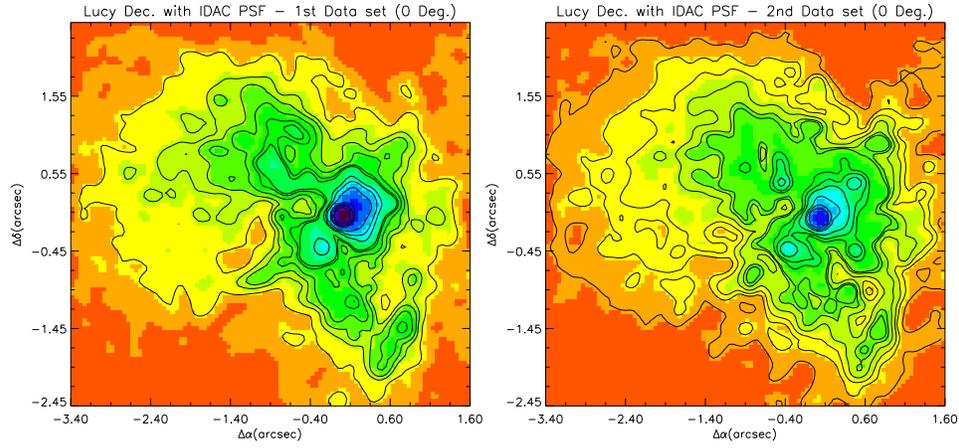}}}
\vspace{-9.5cm}
\caption{K$_c$ data of $\eta$ Car. at 0$^{\circ}$  for the two 
different data sets acquired. 
These images have been obtained after convergence of the blind 
deconvolution algorithm and reconvolution with a Gaussian of 3 
pixels FWHM. This has to be compared with the right hand 
pictures of figure ~\ref{deconv} top and bottom respectively, 
which are the equivalent results for `plucy` deconvolution.}
\label{rec-ima}
\end{figure*}

\begin{figure*}[htb]
\centerline{\resizebox{\hsize}{!}{\includegraphics{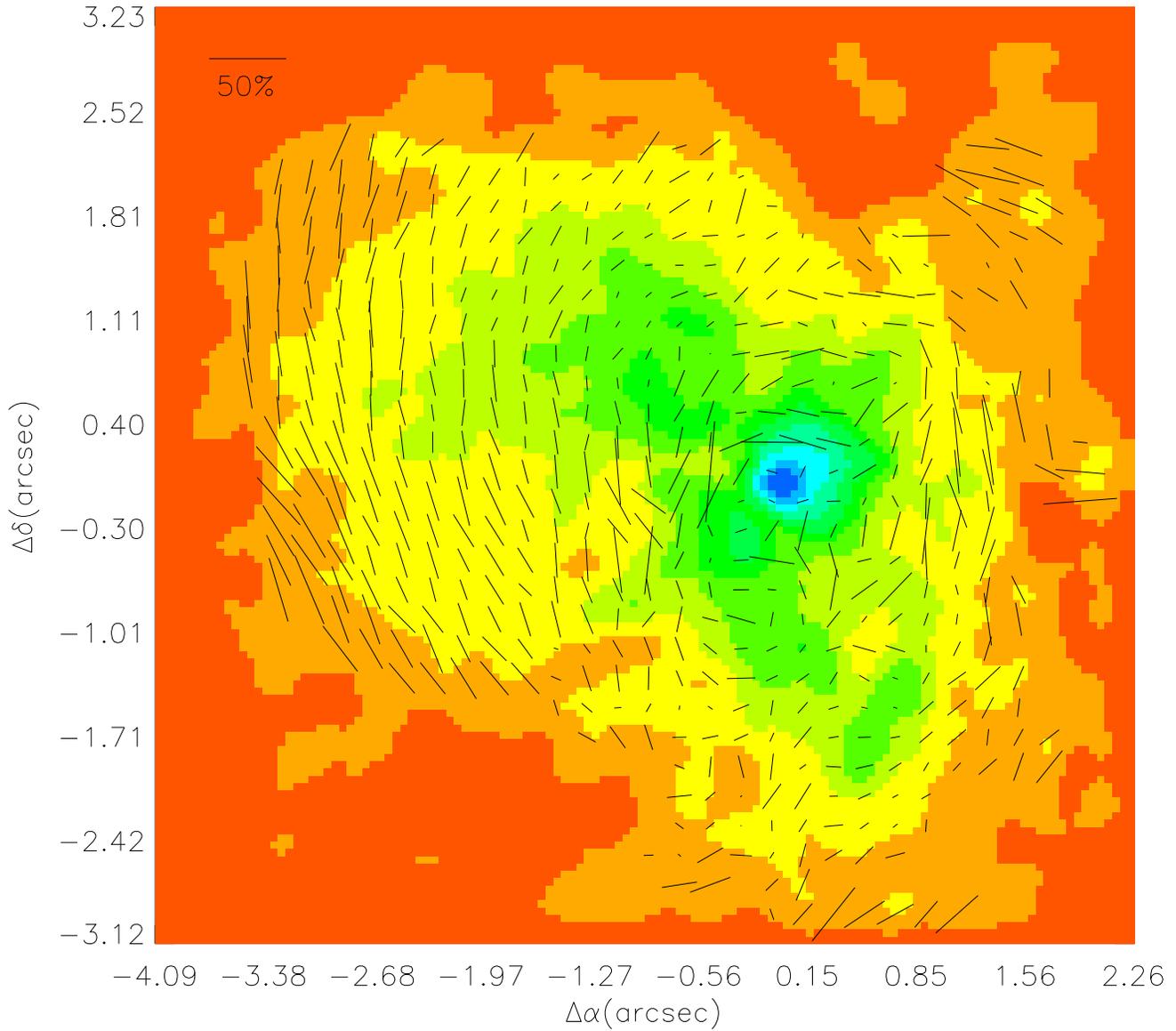}}}
\caption{Polarization map of K$_c$ deconvolved data of $\eta$ Car. 
overplotted on the high resolution intensity map. For clarity, the 
polarization vectors have been calculated on pixels binned 4 by 4 
(0.2$\times$0.2$''$). The intensity map is at the nominal resolution 
of 0.1''}
\label{highp}
\end{figure*}

\begin{figure*}[htb]
\centerline{\resizebox{\hsize}{!}{\includegraphics{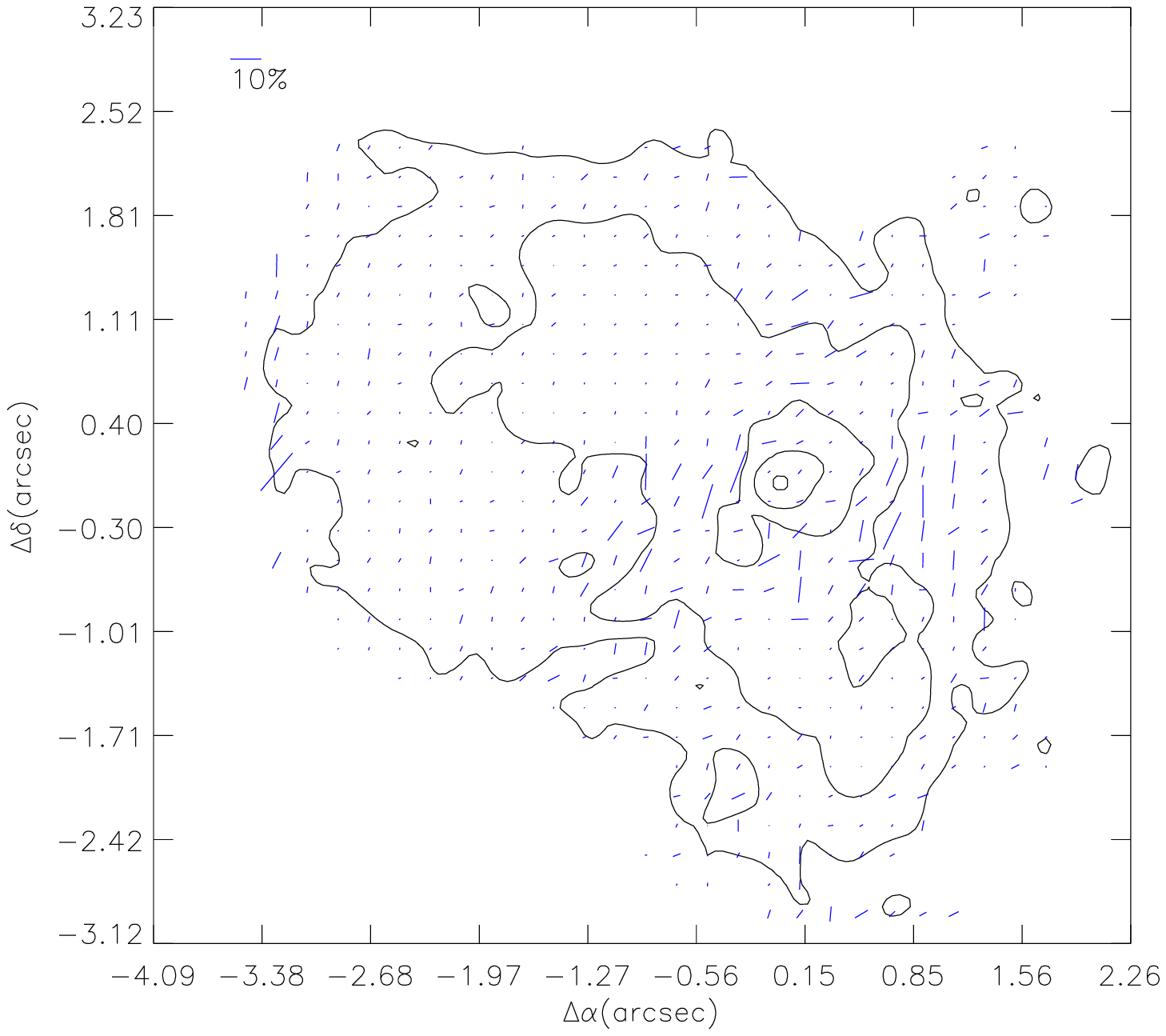}}}
\caption{Differential polarization map of K$_c$ deconvolved data 
of $\eta$ Carinae. It represents the difference between the 
polarization map derived after pure blind deconvolution and 
R-L deconvolution using the PSF derived by blind deconvolution. 
The errors are clearly under 5\% except at the border and thus 
low signal$\_$to$\_$noise level of the nebula and at the position 
of the telescope spider ($\Delta \alpha$ and $\Delta \delta$ = 0). 
Compared to the rest of the map some bigger differences can also 
be found at $\approx$ 0.5'' from $\eta$ Car and is attributed to 
the effect of the PSF wings in the deconvolution process.}
\label{highpcomp}
\end{figure*}


\newpage

\begin{thebibliography}{}
%
\bibitem[1997]{nancy}
Ageorges N., 1997, 'Polarimetric measurements and Deconvolution
Techniques', Proceedings of the NATO-ASI summer school on ``Laser
guide star adaptive optics for Astronomy'', Carg\`ese 29 Sept. - 10
Oct. 1997
%
\bibitem[1997]{agewal}
Ageorges N. \& Walsh J.R., 1997, The Messenger 87, 39
%
\bibitem[1998]{agewalspie}
Ageorges, N., Walsh, J. R., 1998, SPIE Proc. 3353, in press
%
\bibitem[1987]{azzb}
Azzam R. M. A., Bashara, N. M.,1987, Ellipsometry and polarized light,
Elsevier Science, Amsterdam
%
\bibitem[1982]{baihou}
Bailey, J. A., Hough, J. H., 1982, PASP, 94, 618
%
\bibitem[1991]{beich}
Beichman C.A., Ridgway S., 1991, Physics Today, 44, 48
%
\bibitem[1997]{jeanphi}
Berger J.-P., M\'enard F., 1997, Poster proceedings of IAU Symposium 182 on 
Herbig-Haro objects and the birth of low mass stars, F. Malbet \& A. Castet 
eds., 201
%
\bibitem[1969]{bev}
Bevington, P. R., 1969, Data Reduction and Error Analysis for the Physical
Sciences, New York, McGraw-Hill
%
\bibitem[1993]{BH}
Beuzit J.-L. \& Hubin N., 1993, The Messenger 71, 52
%
\bibitem[1997]{beuzit}
Beuzit et al., 1997, Experimental Astronomy 7, 285
%
\bibitem[1997]{christou1}
Christou J. C., Bonaccini D., Ageorges N., 1997, Proc. of SPIE
Vol. 3126, 68
%
\bibitem[1998]{christou2}
Christou J.C., Marchis F., Ageorges N., Bonaccini D. \& Rigaut F., 1998, 
SPIE Proc. 3353, in press
%
\bibitem[1997]{close}
Close L. M. et al., 1997, ApJ 489, 210
%
\bibitem[1995]{code}
Code A.D., Whitney B.A., 1995, ApJ 441, 400
%
\bibitem[1951]{dg}
Davis L. Jr., Greenstein J.L., 1951, ApJ 114, 206 
%
\bibitem[1997]{nico}
Devillard N., 1997, The Messenger 87, 19
%
\bibitem[1984]{draine}
Draine, B. T., Lee, H. M., 1984, ApJ, 285, 89
%
\bibitem[1996]{falcke}
Falcke, H., Davidson, K., Hofmann, K.-H., Weigelt, G., 1996,
A\&A, 306, L17
%
\bibitem[1994]{olaf}
Fischer O., Henning Th., Yorke H.W., 1994, AA 284, 187
%
%
\bibitem[1986]{goodrich}
Goodrich, R. W., 1986, ApJ 311, 882
%
\bibitem[1983]{heckert}
Heckert, P. A., Zeilik, M., 1983, MNRAS, 202, 531
%
\bibitem[1984]{hodapp}
Hodapp K.W., 1984, AA 141, 255
%
%
\bibitem[1995]{hofmann}
Hofmann R., Brandl B., Eckart A., Eisenhauer F., Tacconi-Garman L., 1995, Proc. SPIE 2475, 192
%
\bibitem[1994]{hook}
Hook R.N., Lucy L.B., 1994, Image restorations of high photometric
quality. II. Examples. In Hanisch R.J. \& White R.L. (eds) Space
Telescope Science Institute, The restoration of HST images and spectra II, 86
%
\bibitem[1993]{jeff}
Jefferies S.M., Christou J.C., 1993, ApJ 415, 862
%
\bibitem[1997]{jones}
Jones T.J., 1997, AJ 114, 1393
%
%
\bibitem[1998]{lloyd}
Lloyd-Hart M. et al., 1998, ApJ 493, 950
%
\bibitem[1974]{lucy}
Lucy L., 1974, AJ 79, 745
%
\bibitem[1996]{marco}
Marco, O., Lacombe, F., Bonaccini, D., 1996, ESO Messenger, No. 85, 39
%
\bibitem[1970]{matfor}
Mathewson, D. S., Ford, V. L., 1970, Mem. RAS, 74, 139
%
\bibitem[1990]{mathis}
Mathis, J. S., 1990, ARAA, 28, 37
%
%
\bibitem[1991]{minchin}
Minchin N. R. et al., 1991, MNRAS, 249, 707
%
\bibitem[1993]{nag}
Naghizadeh-Khouei, J., Clarke, D., 1993, A\&A, 274, 968
%
%
\bibitem[1992]{piscalco}
Piirola V., Scaltriti F., Coyne G.V., 1992, Nature Vol 359, No. 6394, 399
%
\bibitem[1972]{richard}
Richardson W.H., 1972, J. Opt. Soc. Am. 62, 55
%
\bibitem[1998]{rigaut}
Rigaut F. et al., 1998, PASP 110, 152
%
%
\bibitem[1997]{sahai}
Sahai R. et al., 1997, ApJ 492, L163
%
\bibitem[1962]{serk}
Serkowski, K., 1962, Adv. in A\&A, ed. Z. Kopal, 1, 289 
%
\bibitem[1991]{sitko}
Sitko M.L. \& Yudong Z., 1991, ApJ 369, 106
%
\bibitem[1995]{shure}
Shure, M., Sellgren K., Jones, T. J., Klebe, D., 1995, AJ, 109, 721
%
\bibitem[1979]{tinbergen}
Tinbergen, J., 1979, A\&ASS, 35, 325
%
\bibitem[1990]{turnshek}
Turnshek D.A. et al., 1990, AJ, 99, 1243
%
\bibitem[1976]{vrba}
Vrba F.J., Strom S.E., Strom K.M., 1976, AJ 81, 958
%
\bibitem[1999]{walage}
Walsh, J. R., Ageorges, N., 1999, A\&A, in preparation
%
\bibitem[1983]{warren}
Warren-Smith, R. F., 1983, MNRAS, 205, 337
%
\bibitem[1986]{weiebe}
Weigelt, G., Ebersberger, J., 1986, A\&A, 163, L5
%
\bibitem[1980]{white}
White, R. L., Schiffer, F. H., Mathis, J. S., 1980, ApJ, 241, 208
%
\bibitem[1992]{whitney}
Whitney B.A., Hartmann L., 1992, ApJ 395, 529
%
\bibitem[1993]{whittet}
Whittet, D. C. B., 1993, Dust in the Galactic Environment,
Bristol, Inst. of Phys.
%
\bibitem[1994]{whitgera}
Whittet, D. C. B. et al., 1994, 
MNRAS 268, 1
%
\bibitem[1977]{witt}
Witt, A. N., 1977, ApJS, 35, 1 
%
\end{thebibliography}
\end{document}